\documentclass[1p,12pt]{elsarticle}

\usepackage{epsfig}

\usepackage{alphalph}

\journal{Astroparticle Physics}

\begin{document}

\begin{frontmatter}

\title{
Measurement of the atmospheric muon flux with a 4~GeV threshold
in the ANTARES neutrino telescope
}

\def\elsauthors{ANTARES collaboration \\ \vspace{0.4cm}}

\author[IFIC]{J.A.~Aguilar}
\author[CPPM]{I.~Al~Samarai}
\author[Colmar]{A.~Albert}
\author[Genova]{M.~Anghinolfi}
\author[Erlangen]{G.~Anton}
\author[IRFU/SEDI]{S.~Anvar}
\author[UPV]{M.~Ardid}
\author[NIKHEF]{A.C.~Assis~Jesus}
\author[NIKHEF]{T.~Astraatmadja\fnref{tag:1}}
\author[CPPM]{J-J.~Aubert}
% ...

\fntext[tag:1]{\scriptsize{Also at University of Leiden, the Netherlands}}

\cortext[cor1]{Corresponding author}

\newpage
\nopagebreak[3]
\address[IFIC]{\scriptsize{IFIC - Instituto de F\'isica Corpuscular, Edificios Investigaci\'on de Paterna, CSIC - Universitat de Val\`encia, Apdo. de Correos 22085, 46071 Valencia, Spain}}\vspace*{0.15cm}
\vspace*{-0.20\baselineskip}
\nopagebreak[3]
\address[CPPM]{\scriptsize{CPPM - Centre de Physique des Particules de Marseille, CNRS/IN2P3 et Universit\'e de la M\'editerran\'ee, 163 Avenue de Luminy, Case 902, 13288 Marseille Cedex 9, France}}\vspace*{0.15cm}
\vspace*{-0.20\baselineskip}
\nopagebreak[3]
\address[Colmar]{\scriptsize{GRPHE - Institut Universitaire de Technologie de Colmar, 34 rue du Grillenbreit BP 50568 - 68008 Colmar, France }}\vspace*{0.15cm}
\vspace*{-0.20\baselineskip}
\nopagebreak[3]
\address[Genova]{\scriptsize{INFN - Sezione di Genova, Via Dodecaneso 33, 16146 Genova, Italy}}\vspace*{0.15cm}
\vspace*{-0.20\baselineskip}
\nopagebreak[3]
\address[Erlangen]{\scriptsize{Friedrich-Alexander-Universit\"{a}t Erlangen-N\"{u}rnberg, Erlangen Centre for Astroparticle Physics, Erwin-Rommel-Str. 1, D-91058 Erlangen, Germany}}\vspace*{0.15cm}
\vspace*{-0.20\baselineskip}
\nopagebreak[3]
\address[IRFU/SEDI]{\scriptsize{Direction des Sciences de la Mati\`ere - Institut de recherche sur les lois fondamentales de l'Univers - Service d'Electronique des D\'etecteurs et d'Informatique, CEA Saclay, 91191 Gif-sur-Yvette Cedex, France}}\vspace*{0.15cm}
\vspace*{-0.20\baselineskip}
\nopagebreak[3]
\address[UPV]{\scriptsize{Institut de Gesti\'o Integrada de Zones Costaneres (IGIC) - Universitat Polit\`ecnica de Val\`encia. Cra. Nazaret-Oliva S/N E-46730 Gandia, Val\`encia, Spain}}\vspace*{0.15cm}
\vspace*{-0.20\baselineskip}
\nopagebreak[3]
\address[NIKHEF]{\scriptsize{FOM Instituut voor Subatomaire Fysica Nikhef, Science Park 105, 1098 XG Amsterdam, The Netherlands}}\vspace*{0.15cm}
\vspace*{-0.20\baselineskip}
\nopagebreak[3]
\address[APC]{\scriptsize{APC - Laboratoire AstroParticule et Cosmologie, UMR 7164 (CNRS, Universit\'e Paris 7 Diderot, CEA, Observatoire de Paris) 10, rue Alice Domon et L\'eonie Duquet 75205 Paris Cedex 13,  France}}\vspace*{0.15cm}
\vspace*{-0.20\baselineskip}
\nopagebreak[3]
\address[LAM]{\scriptsize{LAM - Laboratoire d'Astrophysique de Marseille, P\^ole de l'\'Etoile Site de Ch\^ateau-Gombert, rue Fr\'ed\'eric Joliot-Curie 38,  13388 Marseille cedex 13, France }}\vspace*{0.15cm}
\vspace*{-0.20\baselineskip}
\nopagebreak[3]
\address[Bologna-UNI]{\scriptsize{Dipartimento di Fisica dell'Universit\`a, Viale Berti Pichat 6/2, 40127 Bologna, Italy}}\vspace*{0.15cm}
\vspace*{-0.20\baselineskip}
\nopagebreak[3]
\address[Bologna]{\scriptsize{INFN - Sezione di Bologna, Viale Berti Pichat 6/2, 40127 Bologna, Italy}}\vspace*{0.15cm}
\vspace*{-0.20\baselineskip}
\nopagebreak[3]
\address[Roma-UNI]{\scriptsize{Dipartimento di Fisica dell'Universit\`a "La Sapienza", P.le Aldo Moro 2, 00185 Roma, Italy}}\vspace*{0.15cm}
\vspace*{-0.20\baselineskip}
\nopagebreak[3]
\address[Rome]{\scriptsize{INFN -Sezione di Roma, P.le Aldo Moro 2, 00185 Roma, Italy}}\vspace*{0.15cm}
\vspace*{-0.20\baselineskip}
\nopagebreak[3]
\address[Clermont-Ferrand]{\scriptsize{Laboratoire de Physique Corpusculaire, IN2P3-CNRS, Universit\'e Blaise Pascal, Clermont-Ferrand, France}}\vspace*{0.15cm}
\vspace*{-0.20\baselineskip}
\nopagebreak[3]
\address[Pisa-UNI]{\scriptsize{Dipartimento di Fisica dell'Universit\`a, Largo B. Pontecorvo 3, 56127 Pisa, Italy}}\vspace*{0.15cm}
\vspace*{-0.20\baselineskip}
\nopagebreak[3]
\address[Pisa]{\scriptsize{INFN - Sezione di Pisa, Largo B. Pontecorvo 3, 56127 Pisa, Italy}}\vspace*{0.15cm}
\vspace*{-0.20\baselineskip}
\nopagebreak[3]
\address[INAF]{\scriptsize{INAF-IASF, via P. Gobetti 101, 40129 Bologna, Italy}}\vspace*{0.15cm}
\vspace*{-0.20\baselineskip}
\nopagebreak[3]
\address[GEOAZUR]{\scriptsize{G\'eoazur - Universit\'e de Nice Sophia-Antipolis, CNRS/INSU, IRD, Observatoire de la C\^ote d'Azur and Universit\'e Pierre et Marie Curie, F-06235, BP 48, Villefranche-sur-mer, France}}\vspace*{0.15cm}
\vspace*{-0.20\baselineskip}
\nopagebreak[3]
\address[Bari]{\scriptsize{INFN - Sezione di Bari, Via E. Orabona 4, 70126 Bari, Italy}}\vspace*{0.15cm}
\vspace*{-0.20\baselineskip}
\nopagebreak[3]
\address[LNS]{\scriptsize{INFN - Laboratori Nazionali del Sud (LNS), Via S. Sofia 44, 95123 Catania, Italy}}\vspace*{0.15cm}
\vspace*{-0.20\baselineskip}
\nopagebreak[3]
\address[IRFU/SPP]{\scriptsize{Direction des Sciences de la Mati\`ere - Institut de recherche sur les lois fondamentales de l'Univers - Service de Physique des Particules, CEA Saclay, 91191 Gif-sur-Yvette Cedex, France}}\vspace*{0.15cm}
\vspace*{-0.20\baselineskip}
\nopagebreak[3]
\address[COM]{\scriptsize{COM - Centre d'Oc\'eanologie de Marseille, CNRS/INSU et Universit\'e de la M\'editerran\'ee, 163 Avenue de Luminy, Case 901, 13288 Marseille Cedex 9, France}}\vspace*{0.15cm}
\vspace*{-0.20\baselineskip}
\nopagebreak[3]
\address[UPS]{\scriptsize{Universit\'e Paris-Sud 11 - D\'epartement de Physique - F - 91403 Orsay Cedex, France}}\vspace*{0.15cm}
\vspace*{-0.20\baselineskip}
\nopagebreak[3]
\address[Genova-UNI]{\scriptsize{Dipartimento di Fisica dell'Universit\`a, Via Dodecaneso 33, 16146 Genova, Italy}}\vspace*{0.15cm}
\vspace*{-0.20\baselineskip}
\nopagebreak[3]
\address[NIOZ]{\scriptsize{Royal Netherlands Institute for Sea Research (NIOZ), Landsdiep 4,1797 SZ 't Horntje (Texel), The Netherlands}}\vspace*{0.15cm}
\vspace*{-0.20\baselineskip}
\nopagebreak[3]
\address[KVI]{\scriptsize{Kernfysisch Versneller Instituut (KVI), University of Groningen, Zernikelaan 25, 9747 AA Groningen, The Netherlands}}\vspace*{0.15cm}
\vspace*{-0.20\baselineskip}
\nopagebreak[3]
\address[UU]{\scriptsize{Universiteit Utrecht, Faculteit Betawetenschappen, Princetonplein 5, 3584 CC Utrecht, The Netherlands}}\vspace*{0.15cm}
\vspace*{-0.20\baselineskip}
\nopagebreak[3]
\address[UvA]{\scriptsize{Universteit van Amsterdam, Institut voor Hoge-Energiefysika, Science Park 105, 1098 XG Amsterdam, The Netherlands}}\vspace*{0.15cm}
\vspace*{-0.20\baselineskip}
\nopagebreak[3]
\address[Catania-UNI]{\scriptsize{Dipartimento di Fisica ed Astronomia dell'Universit\`a, Viale Andrea Doria 6, 95125 Catania, Italy}}\vspace*{0.15cm}
\vspace*{-0.20\baselineskip}
\nopagebreak[3]
\address[WIN]{\scriptsize{University of Wisconsin - Madison, 53715, WI, USA}}\vspace*{0.15cm}
\vspace*{-0.20\baselineskip}
\nopagebreak[3]
\address[ISS]{\scriptsize{Institute for Space Sciences, R-77125 Bucharest, M\u{a}gurele, Romania     }}\vspace*{0.15cm}
\vspace*{-0.20\baselineskip}
\nopagebreak[3]
\address[IPHC]{\scriptsize{IPHC-Institut Pluridisciplinaire Hubert Curien, Universit\'e Louis Pasteur (Strasbourg 1) et IN2P3/CNRS, 23 rue du Loess, BP 28, 67037 Strasbourg Cedex 2, France}}\vspace*{0.15cm}
\vspace*{-0.20\baselineskip}
\nopagebreak[3]
\address[ITEP]{\scriptsize{ITEP - Institute for Theoretical and Experimental Physics, B. Cheremushkinskaya 25, 117218 Moscow, Russia}}\vspace*{0.15cm}
\vspace*{-0.20\baselineskip}
\nopagebreak[3]

\begin{abstract}

A new method for the measurement of the muon flux in the deep-sea ANTARES neutrino
telescope and its dependence on the depth is presented.
The method is based on the observation of coincidence signals in adjacent storeys of the detector.
This yields an energy threshold of about 4 GeV.
The main sources of optical background are the decay of $^{40}$K and the bioluminescence in the sea water.
The $^{40}$K background is used to calibrate the efficiency of the photo-multiplier tubes.

\end{abstract}


\end{frontmatter}


\section{Introduction}\label{introduction}
\begin{figure}
\begin{center}
\epsfig{figure=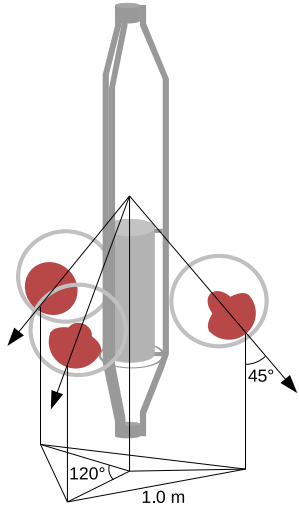,height=2.9in}
\caption{Schematic view of one of the detector storeys.
The storey contains three transparent glass vessels (17" diameter),
each housing a 10" PMT.
The PMTs are oriented downwards
at an angle of 45$^o$ from the vertical and 
with an angle of 120$^o$ from each other.
The distance between the centers of the photo-cathodes 
of the PMTs is 1.0 m.
The storey also contains an electronics module for the readout of the PMTs.
}
\end{center}
\label{detector}
\end{figure}

The ANTARES neutrino telescope is installed in the Mediterranean Sea, 40~km offshore from Toulon (France)
at a depth of 2475 m \cite{Proposal, JCarr}.
The elements of the detector are arranged in twelve vertical lines, each of which is attached to the seabed with
an anchor and kept taut by a buoy.
A detector line comprises 25 storeys distributed along the length
of the 450~m long electro-mechanical cable, starting 100~m above the seabed.
The storeys are separated by 14.5~m and contain a triplet of large-area hemispherical 10" photo-multiplier tubes (PMTs),
arranged as in Figure~\ref{detector} \cite{OM, PMT, DAQ}.
The detector has been operated in partial configurations since March 2006 and was completed in May 2008 \cite{JCarr}.
First results are published in \cite{line1paper, line5paper, PointSourcePaper}.

A large flux of muons produced in the atmosphere by high energy cosmic rays passes through the detector.
The Cherenkov light produced in water by a muon with an energy in excess of 4 GeV can generate correlated signals in adjacent storeys.
Background light, mainly due to $^{40}$K decays and bioluminescent organisms, is also present \cite{BackgroundLight}.
Although a single $^{40}$K decay
will produce a relatively small number of Cherenkov photons,
it can be observed as a coincidence between neighbouring PMTs within a storey
if the decay occurs in the vicinity.
Bioluminescence is a single photon process
and contributes only to the accidental coincidence rate \cite{Biolumninescence, MILOMpaper}.

The present analysis employs a coincidence method to identify $^{40}$K decays and 
low-energy atmospheric muons on a statistical basis.
The flux of atmospheric muons is measured at 24 different depths.
The $^{40}$K signal is used to calibrate the efficiency of the PMTs.

The paper is organized as follows.
The calibration procedure is introduced in Section \ref{k40}.
The method to determine the muon flux is presented in Section \ref{eventrate}.
The depth dependence is discussed in Section \ref{DIR}.

\section{Calibration with potassium-40}\label{k40}
\begin{figure}
\begin{center}
\epsfig{figure=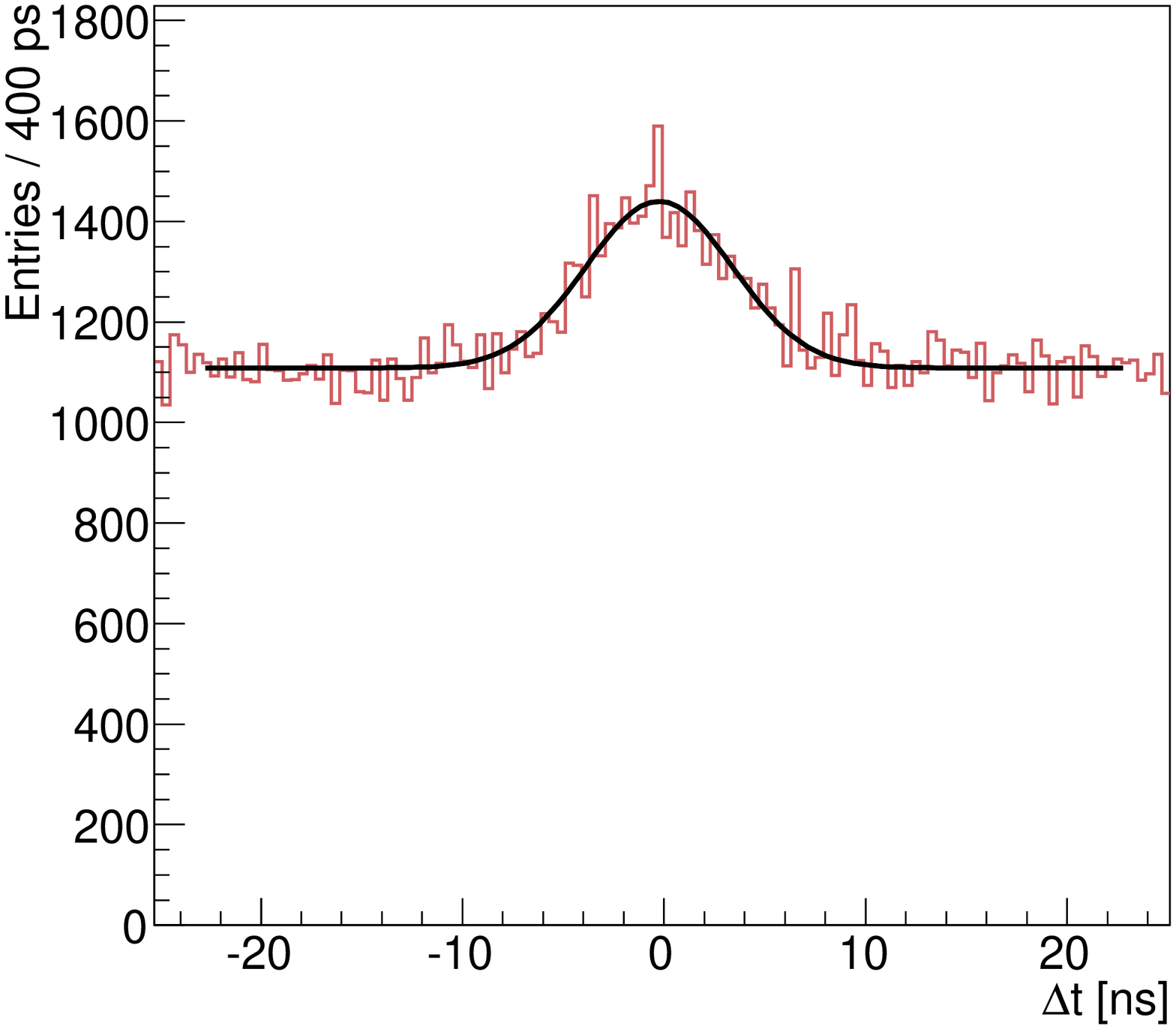,width=2.9in}
\epsfig{figure=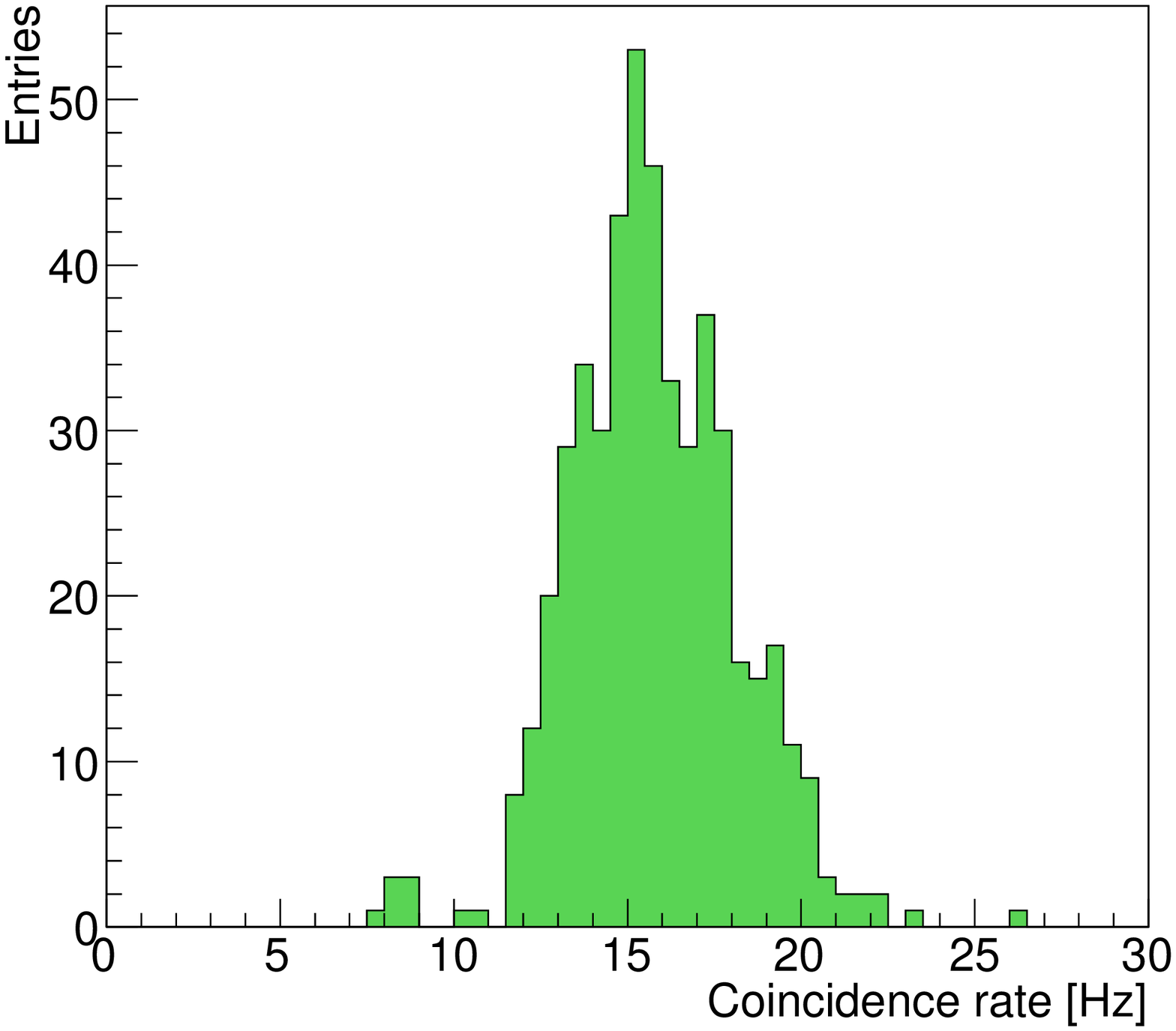,width=2.9in}
\caption{Left: Example distribution of the measured time differences between hits from neighbouring
         PMTs in the same storey. The solid line is a fit to a Gaussian
         peak plus a flat background (see text).
         Right: Histogram of the coincidence rates (background subtracted) observed in the twelve detector lines.}
\end{center}
\label{k40plot}
\end{figure}

Potassium-40 is a radioactive isotope naturally present in the sea water.
The decay~ $^{40}$K $\rightarrow$ e$^-$ $\overline{\nu}_e$ $^{40}$Ca ~yields an electron with an energy up to 1.3 MeV.
This energy exceeds the Cherenkov threshold for electrons in sea water (0.25~MeV), and is sufficient
to produce up to 150 Cherenkov photons.
Another source of electrons above the Cherenkov threshold is the Compton scattering
of 1.46 MeV photons, which are produced in the process of electron capture
$^{40}$K + e$^{-}$ $\rightarrow$ $^{40}$Ar$^{*}$ + ${\nu}_e$, followed by $^{40}$Ar$^{*}$ $\rightarrow$ $^{40}$Ar + $\gamma$.
If the decay occurs within a few meters of a detector storey,
a coincident signal may be recorded by two of the three PMTs on the storey.
This is referred to as a local coincidence.
An example of the distribution of the measured time differences between hits in 
neighbouring PMTs in the same storey is shown in Figure~\ref{k40plot} (left).
A clear peak around 0 ns is visible.
This peak is mainly due to single $^{40}$K decays producing coincident signals.
The data has been fitted to a sum of a Gaussian distribution and a flat background.
The full width half maximum of the Gaussian function is about 9~ns.
This width is mainly due to
the spatial distribution of the $^{40}$K decays around the storey.
The rate of genuine coincidences is given by the integral under the peak
and can be extracted from the distribution by subtracting
the flat background of random coincidences.
The positions of the peaks of the time distributions for different pairs of PMTs in the same storey
are consistent with zero and are used to cross-check the time offsets computed by the time calibration system \cite{OpticalBeacons}.

The spatial uniformity and temporal stability of the rate of $^{40}$K decays is used to calibrate the relative efficiencies of the PMTs.
For every detector storey three coincidence rates, $r_{12}$, $r_{23}$ and $r_{31}$, are measured,
which are related to the efficiencies of the three PMTs,
$s_{1}$, $s_{2}$ and $s_{3}$, where the subscript refers to the position of the PMT 
inside the storey.
The efficiencies are derived from the measured rates
assuming that the rate is proportional to the efficiency of each module,
$r_{ij}\propto s_{i} s_{j}$.
From a combined analysis of the coincidence rates and single rates,
the uncertainty of the extracted efficiencies was estimated to be 5\%.
An average coincidence rate of 16 $\pm$ 2 Hz is observed.
This agrees with the expected value of 19 $\pm$ 3 Hz,
obtained from a dedicated Monte Carlo simulation.
The distribution of observed coincidence rates is shown in Figure~\ref{k40plot} (right).
The spread of the coincidence rates 
corresponds to a 10\% spread in the PMT and front-end electronics efficiencies.
The measured efficiencies have an overall normalisation uncertainty of about 15\%,
originating from the limited knowledge of the quantum efficiency and angular response of the PMT.

\section{Measurement of the atmospheric muon flux}\label{eventrate}

A muon passing through the detector can produce correlated signals in several storeys of the line.
In the following, hits from two adjacent storeys with a local coincidence in each storey are considered,
where the local coincidence is defined as a pair of hits detected by two different PMTs of the same storey
within a $\pm$ 20 ns time window.
The rate of accidental coincidence events between adjacent storeys is about 0.1 Hz.
The data have been recorded during dedicated data taking runs, when bioluminescent activity was low compared to $^{40}$K background.

\begin{figure}
\begin{center}
\epsfig{figure=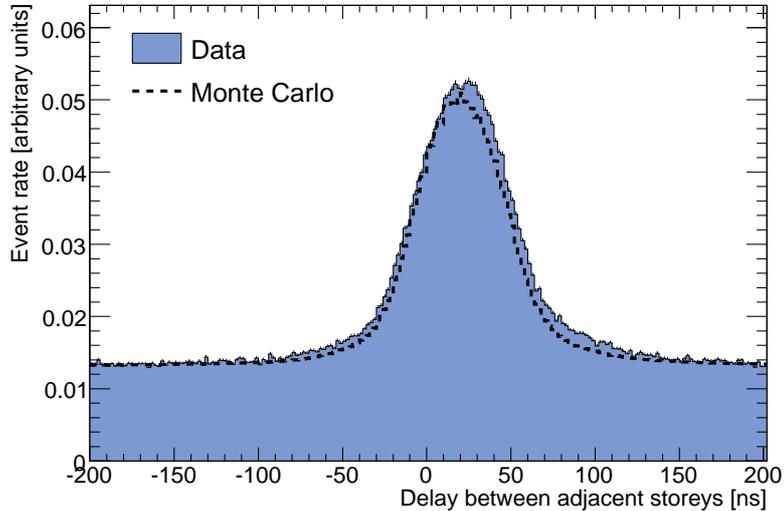,height=2.8in}
\caption{Distribution of the measured time differences between hits from adjacent storeys (lower - upper).
A local coincidence is required in both storeys.
The dashed line corresponds to a Monte Carlo simulation based on MUPAGE (see text).}
\end{center}
\label{a2peak}
\end{figure}

The distribution of the measured time differences between hits in adjacent storeys is presented in Figure~\ref{a2peak}.
A clear peak is visible centred around +20 ns, demonstrating that the majority of muons are downgoing.
The width of the distribution is mainly due to the angular distribution of the atmospheric muons.
The flat background is due to random coincidences from $^{40}$K and bioluminescence.
The tails, which extend beyond $\pm$65 ns, can be attributed to multiple muons and light scattering.
The muon event rate, $R$, is defined as the integral of the peak after subtraction of the flat background.
On average, $R$ is found $\approx$ 0.06 Hz.
Since the probability that a $^{40}$K decay is seen by two PMTs separated by 14.5~m is very small, the $^{40}$K contamination
in $R$ is negligible.

The results of a Monte Carlo simulation based on MUPAGE \cite{MUPAGE}
are also presented in Figure~\ref{a2peak}.
The detector response was simulated using a Geant-4 based model of the PMT angular acceptance
and includes the effect of the detector inefficiencies.
As can be seen from Figure~\ref{a2peak}, the predicted peak is in agreement with the observations.

\section{Depth dependence of the atmospheric muon flux}\label{DIR}

The rate of genuine coincidences between adjacent storeys is mainly due to the flux of vertical and nearly vertical muons.
The muon flux includes, by definition, the corresponding contributions from multiple muons.
The energy threshold for detecting downgoing muons is determined by the minimum track length
to reach two adjacent storeys, and is about 4 GeV.

The efficiency to detect atmospheric muons
can be characterized by a single parameter, $A$,
which gives the ratio between the event rate and the muon flux,
and can be interpreted as an effective area.
From the Monte Carlo simulation based on MUPAGE, the value for a pair of nominal storeys is $A = 87 \, ^{+43}_{-30}$ m$^2$.
The uncertainty originates from the limited knowledge of 
the quantum efficiency and angular acceptance of the PMTs,
as well as uncertainties in the measurements of light absorption length in the sea water \cite{OM, TransmissionOfLight}.
A low energy cutoff of 1 GeV has been used for the flux integration.

The energy spectrum and angular distribution of the muons change with depth.
Therefore, $A$ should be determined as a function of depth.
From the Monte Carlo simulation, the difference between the bottom and top storey is found to be less than 1\%.
This difference has been neglected in the following.

\begin{figure}
\begin{center}
\epsfig{figure=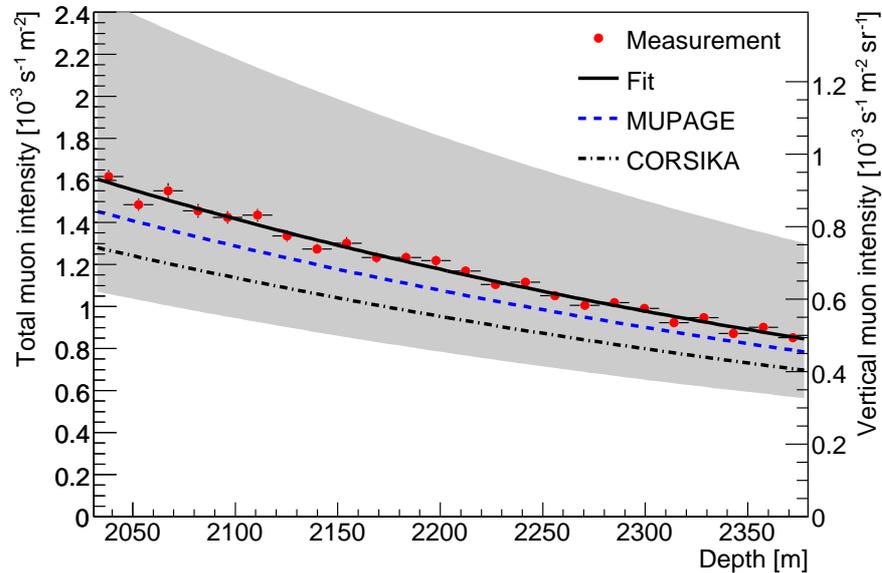,height=3.1in}
\end{center} 
\caption{The measured flux of muons as a function of depth.
The results can be interpreted in terms of total muon intensity (left axis) as well as vertical muon intensity (right axis).
The grey band shows the normalization uncertainty of the data.
The predictions of the Monte Carlo simulations based on MUPAGE and CORSIKA
are shown by dashed and dash-dotted lines, respectively.
}
\label{verticalflux}
\end{figure}

This analysis uses many short calibration runs taken between January 2007 and May 2008.
During that period, in November 2007, the detector was extended from 5 to 10 lines.
The effective live times of the data samples taken with 5 and 10 lines are 4 and 3 hours respectively.
The two subsets of data have been merged together.
The number of active storeys on each line and the efficiency of each storey
is measured following the calibration procedure described above.
From these the average efficiency of
all the storeys at the same depth is computed via Monte Carlo simulation.
The average efficiency for this method is about 58\% with a systematic uncertainty of less than 10\%.
Because a coincidence between two storeys is required,
a measurement of the event rate can be made at 24 different depths.
The event rates are then corrected for the computed
efficiencies and converted into flux units using the effective area $A$.
The depth dependent muon flux thus obtained is shown in Figure~\ref{verticalflux}.

The observed muon flux decreases with increasing depth, $h$, 
as expected from the energy loss of the muon in the water.
An exponential function has been used to fit the data,
$\Phi(h) = \Phi_{0} \times \exp((h_0 - h) / \lambda)$.
We find $\Phi_{0} = 1.71 \pm 0.02 \, \mbox{(stat)} \, ^{+0.92}_{-0.57} \, \mbox{(syst)} \times 10^{-3} \ $m$^{-2}~$s$^{-1}$ at $h_0 = 2000$ m, 
with a slope given by $\lambda = 540 \pm 25$~m.
It is worth noting that the large uncertainty of $A$ only affects the normalization of the measured dependence ($\Phi_{0}$),
but not the slope ($\lambda$).
The uncertainty in $\lambda$ is dominated by the statistical uncertainties of the individual data points and is given by the fit.

The values obtained from the Monte Carlo can be parametrized in the same way. This yields
$\Phi_{0}^\textrm{\scriptsize {MUPAGE}}=1.54 \times 10^{-3} \ $m$^{-2} ~$s$^{-1}$ at 2000 m
and a slope $\lambda^\textrm{\scriptsize {MUPAGE}} = 560 ~$m (Figure~\ref{verticalflux}).
The data are in agreement with the simulation
within a relatively large scale error.
A similar result is obtained from a
Monte Carlo simulation based on CORSIKA \cite{CORSIKA}.
The simulation incorporates the QGSJET model of hadronic interactions and 
the NSU model of the primary cosmic ray spectrum \cite{bugaevflux}.
The propagation of muons in the water was simulated with MUSIC \cite{MUSIC}.
The simulation yields 
$\Phi_{0}^\textrm{\scriptsize {CORSIKA}}=1.36 \times 10^{-3} \ $m$^{-2} ~$s$^{-1}$
and $\lambda^\textrm{\scriptsize {CORSIKA}} = 570 ~$m.
This result, also shown in Figure~\ref{verticalflux},
is compatible with the calculation from MUPAGE.

The muon flux is often presented in terms of vertical muon intensity
\cite{baikal-nt36-1997,amanda-b10-1999,line1paper,line5paper}.
This usually implies a
deconvolution of the measured zenith angle distribution.
The method presented here does not require such a deconvolution procedure and provides
the depth dependence of the muon flux directly.
Hence, the results can be expressed in the form of vertical muon intensity.
For this a constant conversion factor $K$, which translates
the total flux to the corresponding vertical intensity at the same depth, is used.
Using the Monte Carlo simulation, a conversion factor $K = 0.58$~sr$^{-1}$ is found.
The uncertainty of $K$ is small compared to other normalization uncertainties.
The variations with depth are negligible in the depth range of the ANTARES detector.
The result of this conversion is also shown in Figure~\ref{verticalflux}, right vertical scale.
Within uncertainties these results are compatible with the results of \cite{line1paper,line5paper},
which refer to a broader depth range, partly including the depth range covered in this analysis.

\section{Summary}\label{conclusion}
A simple method for the determination of the atmospheric muon flux and its dependence on depth
in the ANTARES neutrino telescope has been presented.
The method is based on the measurement of photon coincidences between adjacent storeys and
has a low energy detection threshold.
The atmospheric muon flux has been measured in the depth range from 2030 to 2380 m with a step of 14.5 m
using a combined data sample of 5 and 10 line detector configurations.
The data have been corrected for the presence of dead channels and unequal efficiencies of the PMTs,
which were measured with a novel calibration technique using the natural radioactivity of sea water.
A reasonable agreement is found between our data and Monte Carlo simulations.

\section*{Acknowledgements}

The authors acknowledge the financial support of the funding agencies:
Centre National de la Recherche Scientifique (CNRS), Commissariat
\`{a} l'Energie Atomique (CEA), Commission Europ\'{e}enne (FEDER fund
and Marie Curie Program), R\'{e}gion Alsace (contrat CPER), R\'{e}gion
Provence-Alpes-C\^{o}te d'Azur, D\'{e}partement du Var and Ville de La
Seyne-sur-Mer, in France; Bundesministerium f\"{u}r Bildung und
Forschung (BMBF), in Germany; Istituto Nazionale di Fisica Nucleare
(INFN), in Italy; Stichting voor Fundamenteel Onderzoek der Materie
(FOM), Nederlandse organisatie voor Wetenschappelijk Onderzoek (NWO),
in the Netherlands; Russian Foundation for Basic Research (RFBR), in
Russia; National Authority for Scientific Research (ANCS) in Romania;
Ministerio de Ciencia e Innovaci\'{o}n (MICINN) and Prometeo of Generalitat Valenciana, in Spain.  We also
acknowledge the technical support of Ifremer, AIM and Foselev Marine
for the sea operation and the CC-IN2P3 for the computing facilities.


\begin{thebibliography}{00}


\bibitem{Proposal} E.~Aslanides {\it et al.},
A deep sea telescope for high energy neutrinos,
arXiv:astro-ph/9907432.

\bibitem{JCarr}
J.~Carr, for the ANTARES collaboration,
First data from the operation of the ANTARES neutrino telescope,
J. Phys. Conf. Ser. {\bf 136} (2008) 022047.

\bibitem{OM} P.~Amram {\it et al.}, 
The ANTARES optical module,
Nucl.\ Instrum.\ Meth.\  A {\bf 484}, 369 (2002).

\bibitem{PMT} J.~A.~Aguilar {\it et al.}, 
Study of Large Hemispherical Photomultiplier Tubes for the ANTARES Neutrino Telescope,
Nucl.\ Instrum.\ Meth.\  A {\bf 555}, 132 (2005).

\bibitem{DAQ} J.~A.~Aguilar {\it et al.}, 
The data acquisition system for the ANTARES neutrino telescope.
Nucl.\ Instrum.\ Meth.\  A {\bf 570}, 107 (2007). 

\bibitem{line1paper} M.~Ageron {\it et al.}, 
Performance of the First ANTARES Detector Line,
Astropart.\ Phys.\  {\bf 31}, 277 (2009).

\bibitem{line5paper} J.~A.~Aguilar {\it et al.}, 
Angular distribution of atmospheric muons measured with the 5-line ANTARES detector,
Preprint.

\bibitem{PointSourcePaper} J.~A.~Aguilar {\it et al.}, 
Search for cosmic neutrino point sources with the 5-line ANTARES telescope,
arXiv:0909.1262 [astro-ph-HE], Submitted to Journal of Physics G.

\bibitem{BackgroundLight} P.~Amram {\it et al.}, 
Background light in potential sites for the ANTARES undersea neutrino telescope,
Astropart.\ Phys.\  {\bf 13}, 127 (2000).

\bibitem{Biolumninescence} P.~M.~S.~Monk,
Physical chemistry: understanding our chemical world, 
John Wiley and Sons Ltd, Chichester, West Sussex, England,
2004,
p. 478.

\bibitem{MILOMpaper} J.~A.~Aguilar {\it et al.}, 
First results of the Instrumentation Line for the deep-sea ANTARES neutrino telescope,
Astropart. Phys. {\bf 26}, 314 (2006). 

\bibitem{OpticalBeacons} M. Ageron {\it et al.}, 
The ANTARES Optical Beacon System,
Nucl.\ Instrum.\ Meth.\ A {\bf 578}, 498 (2007). 

\bibitem{MUPAGE}
G. Carminati, A. Margiotta and M. Spurio,
Atmospheric MUons from PArametric formulas: a fast GEnerator for neutrino telescopes (MUPAGE).
Comput. Phys. Commun. {\bf 179}, 915 (2008).

\bibitem{TransmissionOfLight} J.~A.~Aguilar {\it et al.}, 
Transmission of light in deep sea water at the site of the Antares neutrino telescope,
Astropart.\ Phys. {\bf 23}, 131 (2005). 

\bibitem{CORSIKA} D.~Heck, G.~Schatz, T.~Thouw, J.~Knapp and J.~N.~Capdevielle,
CORSIKA: A Monte Carlo code to simulate extensive air showers,
Forschungszentrum Karlsruhe Report FZKA-6019, 1998.

\bibitem{bugaevflux} S.~I.~Nikolsky, I.~N.~Stamenov and S.~Z.~Ushev,
The composition of cosmic rays at energies of 10 to the 15th eV and higher,
Sov.\ Phys.\ JETP {\bf 60}, 10 (1984)
[Zh.\ Eksp.\ Teor.\ Fiz.\  {\bf 87}, 18 (1984)].

\bibitem{MUSIC} P.~Antonioli, C.~Ghetti, E.~V.~Korolkova, V.~A.~Kudryavtsev and G.~Sartorelli,
A three-dimensional code for muon propagation through the rock: MUSIC,
Astropart.\ Phys.\  {\bf 7}, 357 (1997). 

\bibitem{baikal-nt36-1997}
I.~A.~Belolaptikov {\it et al.}, 
The Baikal underwater neutrino telescope: Design, performance, and first results,
Astropart.\ Phys.\  {\bf 7}, 263 (1997).

\bibitem{amanda-b10-1999} E.~Andres {\it et al.}, 
The AMANDA neutrino telescope: Principle of operation and first results,
Astropart.\ Phys.\  {\bf 13}, 1 (2000).

\end{thebibliography}
\end{document}